\documentclass[preprint,prd,aps,showpacs,showkeys,nofootinbib]{revtex4}
\usepackage{graphicx}
\usepackage{dcolumn}
\usepackage{bm}
\begin{document}
\title{The one loop contributions to c(t) electric
dipole moment in the CP violating BLMSSM}
\author{Shu-Min Zhao$^1$\footnote{zhaosm@hbu.edu.cn}, Tai-Fu Feng$^{1}$\footnote{fengtf@hbu.edu.cn}, Zhong-Jun Yang$^{1}$,
Hai-Bin Zhang$^{1}$, Xing-Xing Dong$^1$,  Tao-Guo$^2$}
\affiliation{$^1$ Department of Physics, Hebei University, Baoding 071002,
China\\
$^2$ School of Mathematics and Science, Hebei University of Geosciences, Shijiazhuang 050031, China}
\date{\today}

\begin{abstract}
In the CP violating supersymmetric extension of the standard model with local gauged
baryon and lepton symmetries(BLMSSM), there are new CP violating sources which can
give new contributions to the quark electric dipole
moment (EDM). Considering the CP violating phases, we analyze the EDMs of the quarks c and t.
We take into account the contributions from the one loop diagrams.
The numerical results are analyzed with some assumptions on the relevant parameter space.
The numerical results for the c and t EDMs can reach large values.
\end{abstract}

\pacs{\emph{11.30.Er, 12.60.Jv,14.80.Cp}}

\keywords{CP violating,  electric dipole moment, local gauge symmetry}

\maketitle

\section{Introduction}
 The CP violation found in the K- and B system \cite{CPVKB} can be well explained in
the standard model. It is well known that, the electric dipole moment(EDM)
of elementary particle is a clear sinal of CP violation\cite{cp1}. The Cabbibo-Kobayashi-Maskawa(CKM)
phase is the only source of CP violation in the SM, which has ignorable effect on the EDM of
elementary particle. In the SM, even to two loop order, the EDM of a fermion does not appear, and there
are partial cancelation between the three loop contributions\cite{SMTwoLoop}. If EDM of an elementary fermion is
detected, one can confirm there are new CP phases and  physics beyond the SM.

Though SM has obtained large successes with the detection of the lightest CP-even Higgs $h^0$\cite{Higgs},
it is unable to explain some phenomena. Physicists consider the SM should be a low energy
effective theory of a large model. The minimal supersymmetric extension of the standard
model(MSSM) is very favorite and people have been interested in it for a long time\cite{MSSM}. There are also many
new models beyond the SM, such as $\mu\nu$SSM\cite{munuSSM}. Generally speaking, the new models introduce
new CP-violating phases that can affect the EDMs of fermions, $B^0-\bar{B}^0$ mixing et al.
The EDMs of electron and neutron are strict constraints on the CP-violating phases\cite{EDM}.
In the models beyond SM, there are new CP-violating phases which can give large contributions
to electron and neutron EDMs\cite{NEEDM}. To make the MSSM predictions of electron and neutron EDMs
under the experiment upper bounds, there are three possibilities\cite{xiangxiao}:
1 the CP-violating phases are very small, 2 varies contributions cancel with each other in
some special parameter spaces, 3 the supersymmetry particles are very heavy at several TeV order.

Taking into account the local gauged $B$ and $L$, people obtain
the minimal supersymmetric extension of the SM, which is the so called BLMSSM\cite{BLMSSM1}.
The authors in Ref.\cite{BLMSSM2} first proposed BLMSSM, where they studied some phenomena.
At TeV scale, the local gauge symmetries of BLMSSM breaks spontaneously. Therefore, in BLMSSM R-parity is violated
and the asymmetry of matter-antimatter in the universe can be explained.
We have studied the lightest CP-even Higgs mass and  the
 decays $h^0\rightarrow VV, ~V=(\gamma,Z, W)$\cite{weBLMSSM} in the BLMSSM, where
 some other processes\cite{weBLNCP} are also researched. Taking the CP-violating phases with nonzero values,
 the neutron EDM, lepton EDM and $B^0-\bar{B}^0$ mixing are researched in this model\cite{weBLCPV}.

From neutron experimental data, the bounding of top EDM is analyzed\cite{TopEDM}.
Taking into account the precise measurements of the electron and neutron EDMs,
the upper limits of heavy quark EDMs are also discussed\cite{HQEDM}.
The upper limits on the EDMs of heavy quarks are researched from $e^+e^-$ annihilation\cite{HQEEEDM}.
In the CP-violating MSSM,
the authors study c quark EDM including two loop gluino contributions\cite{BFEDM}. There are also other works
on the c quark EDM\cite{OCEDM}. Considering the pre-existing works, the upper bounds of EDMs for c and t are about
$d_c<5.0\times 10^{-17} e.cm$ and $d_t<3.06\times 10^{-15} e.cm$.
In this work, we calculate the EDMs of charm quark and top quark in the framework of the CP-violating  BLMSSM.
At low energy scale, the quark chromoelectric dipole moment(CEDM) can give important
contributions to the quark EDM. So, we also study the quark CEDM with the renormalization group
equations.

In Section 2, we briefly introduce the BLMSSM and show the needed mass matrices and couplings, after this introduction.
The EDMs(CEDMs) of c and t are researched in Section 3.
In Section 4, we give out the input parameters and calculate the numerical results.  The last Section is used to discuss
the results and the allowed parameter space.

\section{The BLMSSM}

\indent\indent
Considering the local gauge symmetries of B(L) and enlarging the local gauge group of the SM to
$SU(3)_{C}\otimes SU(2)_{L}\otimes U(1)_{Y}\otimes U(1)_{B}\otimes U(1)_{L}$  one can obtain the BLMSSM model\cite{BLMSSM1}.
In the BLMSSM, there are the exotic superfields
including the new quarks $\hat{Q}_{4}\sim(3,\;2,\;1/6,\;B_{4},\;0)$,
$\hat{U}_{4}^c\sim(\bar{3},\;1,\;-2/3,\;-B_{4},\;0)$,
$\hat{D}_{4}^c\sim(\bar{3},\;1,\;1/3,\;-B_{4},\;0)$,
$\hat{Q}_{5}^c\sim(\bar{3},\;2,\;-1/6,\;-(1+B_{4}),\;0)$, $\hat{U}_{5}\sim(3,\;1,\;2/3,\;1+B_{4},\;0)$,
$\hat{D}_{5}\sim(3,\;1,\;-1/3,\;1+B_{4},\;0)$,
and the new leptons $\hat{L}_{4}\sim(1,\;2,\;-1/2,\;0,\;L_{4})$,
$\hat{E}_{4}^c\sim(1,\;1,\;1,\;0,\;-L_{4})$, $\hat{N}_{4}^c\sim(1,\;1,\;0,\;0,\;-L_{4})$,
$\hat{L}_{5}^c\sim(1,\;2,\;1/2,\;0,\;-(3+L_{4}))$, $\hat{E}_{5}\sim(1,\;1,\;-1,\;0,\;3+L_{4})$,
$\hat{N}_{5}\sim(1,\;1,\;0,\;0,\;3+L_{4})$ to cancel the $B$ and $L$ anomalies.

With the detection of the lightest CP even Higgs $h^0$ at LHC\cite{Higgs},
Higgs mechanism is very convincing for particle physics, and BLMSSM is based on the Higgs mechanism.
The introduced Higgs superfields $\hat{\Phi}_{L}(1,\; 1,\; 0, \;0,\; -2),\;\hat{\varphi}_{L}(1, \;1, \;0,\; 0,\; 2)$  and
$\hat{\Phi}_{B}(1, \;1, \;0, \;1,\; 0),\;\hat{\varphi}_{B}(1,\; 1,\; 0, \;-1, \;0)$ break lepton number and baryon number spontaneously.
These Higgs superfields acquire nonzero vacuum expectation values (VEVs)
and provide masses to the exotic leptons and exotic quarks.
To make the heavy exotic quarks unstable
 the superfields $\hat{X}(1,\; 1,\; 0,\; 2/3+B_4,\; 0)$,
$\hat{X}^\prime(1,\; 1,\; 0,\;-(2/3+B_4),\; 0)$ are introduced in the BLMSSM.

The $SU(2)_L$ doublets $H_{u},\;H_{d}$ obtain nonzero VEVs $\upsilon_{u},\;\upsilon_{d}$,
\begin{eqnarray}
&&H_{u}=\left(\begin{array}{c}H_{u}^+\\{1\over\sqrt{2}}\Big(\upsilon_{u}+H_{u}^0+iP_{u}^0\Big)\end{array}\right)\;,~~~~
H_{d}=\left(\begin{array}{c}{1\over\sqrt{2}}\Big(\upsilon_{d}+H_{d}^0+iP_{d}^0\Big)\\H_{d}^-\end{array}\right)\;.
\end{eqnarray}
 The $SU(2)_L$ singlets $\Phi_{B},\;\varphi_{B}$ and $\Phi_{L},\;
\varphi_{L}$ obtain nonzero VEVs
 $\upsilon_{{B}},\;\overline{\upsilon}_{{B}}$ and $\upsilon_{L},\;\overline{\upsilon}_{L}$ respectively,
 \begin{eqnarray}
&&\Phi_{B}={1\over\sqrt{2}}\Big(\upsilon_{B}+\Phi_{B}^0+iP_{B}^0\Big)\;,~~~~~~~~~
\varphi_{B}={1\over\sqrt{2}}\Big(\overline{\upsilon}_{B}+\varphi_{B}^0+i\overline{P}_{B}^0\Big)\;.
\nonumber\\
&&\Phi_{L}={1\over\sqrt{2}}\Big(\upsilon_{L}+\Phi_{L}^0+iP_{L}^0\Big)\;,~~~~~~~~~~
\varphi_{L}={1\over\sqrt{2}}\Big(\overline{\upsilon}_{L}+\varphi_{L}^0+i\overline{P}_{L}^0\Big)\;.
\label{VEVs}
\end{eqnarray}
Therefore, the local gauge symmetry $SU(2)_{L}\otimes U(1)_{Y}\otimes U(1)_{B}\otimes U(1)_{L}$
breaks down to the electromagnetic symmetry $U(1)_{e}$.

We show the superpotential of BLMSSM \cite{weBLMSSM}
\begin{eqnarray}
&&{\cal W}_{{BLMSSM}}={\cal W}_{{MSSM}}+{\cal W}_{B}+{\cal W}_{L}+{\cal W}_{X}\;,
\label{superpotential1}
\nonumber\\&&{\cal W}_{B}=\lambda_{Q}\hat{Q}_{4}\hat{Q}_{5}^c\hat{\Phi}_{B}+\lambda_{U}\hat{U}_{4}^c\hat{U}_{5}
\hat{\varphi}_{B}+\lambda_{D}\hat{D}_{4}^c\hat{D}_{5}\hat{\varphi}_{B}+\mu_{B}\hat{\Phi}_{B}\hat{\varphi}_{B}
\nonumber\\
&&\hspace{1.2cm}
+Y_{{u_4}}\hat{Q}_{4}\hat{H}_{u}\hat{U}_{4}^c+Y_{{d_4}}\hat{Q}_{4}\hat{H}_{d}\hat{D}_{4}^c
+Y_{{u_5}}\hat{Q}_{5}^c\hat{H}_{d}\hat{U}_{5}+Y_{{d_5}}\hat{Q}_{5}^c\hat{H}_{u}\hat{D}_{5}\;,
\nonumber\\
&&{\cal W}_{L}=Y_{{e_4}}\hat{L}_{4}\hat{H}_{d}\hat{E}_{4}^c+Y_{{\nu_4}}\hat{L}_{4}\hat{H}_{u}\hat{N}_{4}^c
+Y_{{e_5}}\hat{L}_{5}^c\hat{H}_{u}\hat{E}_{5}+Y_{{\nu_5}}\hat{L}_{5}^c\hat{H}_{d}\hat{N}_{5}
\nonumber\\
&&\hspace{1.2cm}
+Y_{\nu}\hat{L}\hat{H}_{u}\hat{N}^c+\lambda_{{N^c}}\hat{N}^c\hat{N}^c\hat{\varphi}_{L}
+\mu_{L}\hat{\Phi}_{L}\hat{\varphi}_{L}\;,
\nonumber\\
&&{\cal W}_{X}=\lambda_1\hat{Q}\hat{Q}_{5}^c\hat{X}+\lambda_2\hat{U}^c\hat{U}_{5}\hat{X}^\prime
+\lambda_3\hat{D}^c\hat{D}_{5}\hat{X}^\prime+\mu_{X}\hat{X}\hat{X}^\prime,
\label{superpotential-BL}
\end{eqnarray}
with ${\cal W}_{{MSSM}}$ representing the superpotential of the MSSM.
The soft breaking terms $\mathcal{L}_{{soft}}$ of the BLMSSM are collected here\cite{BLMSSM1, weBLMSSM}.
\begin{eqnarray}
&&{\cal L}_{{soft}}={\cal L}_{{soft}}^{MSSM}-(m_{{\tilde{N}^c}}^2)_{{IJ}}\tilde{N}_I^{c*}\tilde{N}_J^c
-m_{{\tilde{Q}_4}}^2\tilde{Q}_{4}^\dagger\tilde{Q}_{4}-m_{{\tilde{U}_4}}^2\tilde{U}_{4}^{c*}\tilde{U}_{4}^c
-m_{{\tilde{D}_4}}^2\tilde{D}_{4}^{c*}\tilde{D}_{4}^c
\nonumber\\
&&\hspace{1.3cm}
-m_{{\tilde{Q}_5}}^2\tilde{Q}_{5}^{c\dagger}\tilde{Q}_{5}^c-m_{{\tilde{U}_5}}^2\tilde{U}_{5}^*\tilde{U}_{5}
-m_{{\tilde{D}_5}}^2\tilde{D}_{5}^*\tilde{D}_{5}-m_{{\tilde{L}_4}}^2\tilde{L}_{4}^\dagger\tilde{L}_{4}
-m_{{\tilde{\nu}_4}}^2\tilde{N}_{4}^{c*}\tilde{N}_{4}^c
\nonumber\\
&&\hspace{1.3cm}
-m_{{\tilde{e}_4}}^2\tilde{E}_{_4}^{c*}\tilde{E}_{4}^c-m_{{\tilde{L}_5}}^2\tilde{L}_{5}^{c\dagger}\tilde{L}_{5}^c
-m_{{\tilde{\nu}_5}}^2\tilde{N}_{5}^*\tilde{N}_{5}-m_{{\tilde{e}_5}}^2\tilde{E}_{5}^*\tilde{E}_{5}
-m_{{\Phi_{B}}}^2\Phi_{B}^*\Phi_{B}
\nonumber\\
&&\hspace{1.3cm}
-m_{{\varphi_{B}}}^2\varphi_{B}^*\varphi_{B}-m_{{\Phi_{L}}}^2\Phi_{L}^*\Phi_{L}
-m_{{\varphi_{L}}}^2\varphi_{L}^*\varphi_{L}-\Big(m_{B}\lambda_{B}\lambda_{B}
+m_{L}\lambda_{L}\lambda_{L}+h.c.\Big)
\nonumber\\
&&\hspace{1.3cm}
+\Big\{A_{{u_4}}Y_{{u_4}}\tilde{Q}_{4}H_{u}\tilde{U}_{4}^c+A_{{d_4}}Y_{{d_4}}\tilde{Q}_{4}H_{d}\tilde{D}_{4}^c
+A_{{u_5}}Y_{{u_5}}\tilde{Q}_{5}^cH_{d}\tilde{U}_{5}+A_{{d_5}}Y_{{d_5}}\tilde{Q}_{5}^cH_{u}\tilde{D}_{5}
\nonumber\\
&&\hspace{1.3cm}
+A_{{BQ}}\lambda_{Q}\tilde{Q}_{4}\tilde{Q}_{5}^c\Phi_{B}+A_{{BU}}\lambda_{U}\tilde{U}_{4}^c\tilde{U}_{5}\varphi_{B}
+A_{{BD}}\lambda_{D}\tilde{D}_{4}^c\tilde{D}_{5}\varphi_{B}+B_{B}\mu_{B}\Phi_{B}\varphi_{B}
+h.c.\Big\}
\nonumber\\
&&\hspace{1.3cm}
+\Big\{A_{{e_4}}Y_{{e_4}}\tilde{L}_{4}H_{d}\tilde{E}_{4}^c+A_{{\nu_4}}Y_{{\nu_4}}\tilde{L}_{4}H_{u}\tilde{N}_{4}^c
+A_{{e_5}}Y_{{e_5}}\tilde{L}_{5}^cH_{u}\tilde{E}_{5}+A_{{\nu_5}}Y_{{\nu_5}}\tilde{L}_{5}^cH_{d}\tilde{N}_{5}
\nonumber\\
&&\hspace{1.3cm}
+A_{N}Y_{\nu}\tilde{L}H_{u}\tilde{N}^c+A_{{N^c}}\lambda_{{N^c}}\tilde{N}^c\tilde{N}^c\varphi_{L}
+B_{L}\mu_{L}\Phi_{L}\varphi_{L}+h.c.\Big\}
\nonumber\\
&&\hspace{1.3cm}
+\Big\{A_1\lambda_1\tilde{Q}\tilde{Q}_{5}^cX+A_2\lambda_2\tilde{U}^c\tilde{U}_{5}X^\prime
+A_3\lambda_3\tilde{D}^c\tilde{D}_{5}X^\prime+B_{X}\mu_{X}XX^\prime+h.c.\Big\}\;.
\label{soft-breaking}
\end{eqnarray}
${\cal L}_{{soft}}^{MSSM}$ are the soft breaking terms of MSSM.

\subsection{mass matrix}

From the soft breaking terms and the scalar potential, we deduce the mass squared matrix for superfields X.
\begin{eqnarray}
&&-\mathcal{L}_{X}=(X^*~~~X')\left(     \begin{array}{cc}
  |\mu_{X}|^2+S_{X} &-\mu_{_X}^*B_{X}^* \\
    -\mu_{X}B_{X} & |\mu_{X}|^2-S_{X}\\
    \end{array}\right)  \left( \begin{array}{c}
  X \\  X'^*\\
    \end{array}\right),
    \nonumber\\&&
    S_{X}=\frac{g_{B}^2}{2}(\frac{2}{3}+B_{4})(v_{B}^2-\bar{v}_{B}^2).
   \end{eqnarray}
  We diagonalize the mass squared matrix for the superfields X
 through the unitary transformation,
   \begin{eqnarray}
   ~~~\left(     \begin{array}{c}
  X_{1} \\  X_{2}\\
    \end{array}\right) =Z_{X}^{\dag}\left( \begin{array}{c}
  X \\  X'^*\\
    \end{array}\right),
~~~~~~
Z^{\dag}_{X}\left(  \begin{array}{cc}
  |\mu_{X}|^2+S_{X} &-\mu_{X}^*B_{X}^* \\
    -\mu_{X}B_{X} & |\mu_{X}|^2-S_{X}\\
    \end{array}\right)  Z_{X}=\left(     \begin{array}{cc}
 m_{X_1}^2 &0 \\
    0 & m_{X_2}^2\\
    \end{array}\right). \label{Xchang}
   \end{eqnarray}

 $\psi_X$ and $\psi_{X'}$ are the superpartners of the scalar superfields $X$ and $X'$.
$\psi_X$ and $\psi_{X'}$ can composite four-component Dirac spinors, whose mass term are given out\cite{weBLCPV}
  \begin{eqnarray}
  &&-\mathcal{L}_{\tilde{X}}^{mass}=\mu_X\bar{\tilde{X}}\tilde{X}, ~~~~~~
  \tilde{X}= \left(\begin{array}{c}
  \psi_X\\ \bar{\psi}_{X'}
    \end{array}\right),
  \end{eqnarray}
with $\mu_X$ denoting the mass of $\tilde{X}$.

In the BLMSSM, there are the new baryon boson, the $SU(2)_L$ singlets $\Phi_B$ and $\varphi_B$. Their
superpartners are respectively $\lambda_B$, $\psi_{\Phi_B}$ and $\psi_{\varphi_B}$, and they mix together
producing 3 baryon neutralinos. In the base $(i\lambda_B,\psi_{\Phi_B},\psi_{\varphi_B})$,
the mass mixing matrix $M_{BN}$ is obtained and diagonalized by the rotation matrix $Z_{N_B}$\cite{DCPC}.

  \begin{eqnarray}
&&M_{BN}=\left(  \begin{array}{ccc}
  2m_B &-v_Bg_B & \bar{v}_Bg_B\\-
   v_Bg_B & 0 &-\mu_B\\ \bar{v}_Bg_B&-\mu_B &0
    \end{array}\right),~~~~~ \chi^0_{B_i}= \left(\begin{array}{c}
 k_{B_i}^0\\ \bar{k}_{B_i}^0
    \end{array}\right),
     \nonumber\\&& i\lambda_B=Z_{N_B}^{1i}k_{B_i}^0,~~~~~~
   \psi_{\Phi_B}=Z_{N_B}^{2i}k_{B_i}^0,~~~~~~\psi_{\varphi_B}=Z_{N_B}^{3i}k_{B_i}^0.
   \end{eqnarray}
  $\chi^0_{B_i} (i=1,2,3)$ represent the mass eigenstates of baryon neutralinos.

  The exotic quarks with charged 2/3 is in four-component Dirac spinors,
  whose mass matrix reads as\cite{weBLMSSM}
\begin{equation}
-\mathcal{L}^{mass}_{t'}=(\bar{t}'_{4R}~~~\bar{t}'_{5R})\left(     \begin{array}{cc}
  \frac{1}{\sqrt{2}}\lambda_Qv_B &-\frac{1}{\sqrt{2}}Y_{u_5}v_d \\
    -\frac{1}{\sqrt{2}}Y_{u_4}v_u & \frac{1}{\sqrt{2}}\lambda_u\bar{v}_B\\
    \end{array}\right)  \left( \begin{array}{c}
 t'_{4L} \\   t'_{5L}\\
    \end{array}\right),
   \end{equation}

Using the unitary transformations, the two mass eigenstates of exotic quarks with charged 2/3
are obtained by the rotation matrices $U_{t}$ and $W_{t}$,
\begin{eqnarray}
&&\left( \begin{array}{c}
 t_{4L} \\   t_{5L}\\
    \end{array}\right)=U_{t}^{\dag}\left( \begin{array}{c}
 t'_{4L} \\   t'_{5L}\\
    \end{array}\right),~~~~~~~~\left( \begin{array}{c}
 t_{4R} \\   t_{5R}\\
    \end{array}\right)=W_{t}^{\dag}\left( \begin{array}{c}
 t'_{4R} \\   t'_{5R}\\
    \end{array}\right),
\nonumber\\&&
W_{t}^{\dag} \left(\begin{array}{cc}
  \frac{1}{\sqrt{2}}\lambda_Qv_B &-\frac{1}{\sqrt{2}}Y_{u_5}v_d \\
    -\frac{1}{\sqrt{2}}Y_{u_4}v_u & \frac{1}{\sqrt{2}}\lambda_u\bar{v}_B\\
    \end{array}\right)U_{t}=\texttt{diag}(m_{t_4},m_{t_5}). \label{ExoticQ}
\end{eqnarray}

The mass squared matrix for charged 2/3 exotic squarks $\mathcal{M}^2_{\tilde{t}'}$ is obtained in our previous work\cite{weBLMSSM}.
For saving space in the work, we do not show it here. $\mathcal{M}^2_{\tilde{t}'}$ is diagonalized by
$Z_{\tilde{t}'}$ through the formula $Z_{\tilde{t}'}^\dag\mathcal{M}^2_{\tilde{t}'}Z_{\tilde{t}'}=\texttt{diag}(m^2_{\tilde{\mathcal{U}}_1},
m^2_{\tilde{\mathcal{U}}_2},m^2_{\tilde{\mathcal{U}}_3},m^2_{\tilde{\mathcal{U}}_4})$.

\subsection{needed couplings}

To study quark EDMs, the couplings between photon (gluon) and exotic quarks(exotic squarks)
are necessary.
We derive the couplings between photon (gluon) and exotic quarks.
\begin{eqnarray}
&&\mathcal{L}_{\gamma (g) q'q'}=-\frac{2e}{3}\sum_{i=1}^2\bar{t}_{_{i+3}}\gamma^{\mu}t_{_{i+3}}F_{\mu}
-g_{_3}\sum_{i=1}^2\bar{t}_{_{i+3}}T^a\gamma^{\mu}t_{_{i+3}}G^a_{\mu},
\end{eqnarray}
with $F_{\mu}$ and $G^a_{\mu}$ representing electromagnetic field and gluon field
respectively. $T^a\;(a=1,\;\cdots,\;8)$ are the strong $SU(3)$ gauge group generators.
Similarly,  the couplings between photon (gluon) and exotic squarks are also deduced
\begin{eqnarray}
\mathcal{L}_{\gamma (g) \tilde{q}'\tilde{q}'}=-\frac{2}{3}e\sum_{j,\beta=1}^4\delta_{j\beta}F_{\mu}
\tilde{\mathcal{U}}_{j}^*i\tilde{\partial}^{\mu}
\tilde{\mathcal{U}}_{\beta}-g_3T^a\sum_{j,\beta=1}^4\delta_{j\beta}G^a_{\mu}
\tilde{\mathcal{U}}_{j}^*i\tilde{\partial}^{\mu}
\tilde{\mathcal{U}}_{\beta}.
\end{eqnarray}
From the superpotential $W_X$, one can find there are interactions at tree level for
quark, exotic quark and X. The needed Yukawa interactions can be deduced from the superpotential $W_X$.
The couplings of quark-exotic quark-X are shown in the mass basis,
\begin{eqnarray}
&&\mathcal{L}_{_{Xt'u}}=\sum_{i,j=1}^2\Big((\mathcal{N}^L_{_{t'}})_{_{ij}}X_{_j}\bar{t}_{_{i+3}}P_{_L}u^I
+(\mathcal{N}^R_{_{t'}})_{_{ij}}X_{_j}\bar{t}_{_{i+3}}P_{_R}u^I\Big)+h.c.\nonumber\\
  &&(\mathcal{N}^L_{_{t'}})_{_{ij}}=-\lambda_{_1}(W_t^{\dag})_{_{i1}}(Z_{_X})_{_{1j}},~~~~~~~~ (\mathcal{N}^R_{_{t'}})_{_{ij}}=-\lambda_{_2}^*(U_{_t}^{\dag})_{_{i2}}(Z_{_X})_{_{2j}}.\label{Xt'u}
   \end{eqnarray}
From the superpotential $W_X$, in the same way we can
also obtain another type Yukawa couplings (quark-exotic squark-$\tilde{X}$)\cite{weBLCPV}.
\begin{eqnarray}
\mathcal{L}_{\bar{u}\tilde{X}\tilde{\mathcal{U}}}=-\sum_{i=1}^4\bar{u}\Big(
\lambda_1(Z_{\tilde{t}'})^*_{3i}P_L+\lambda_2(Z_{\tilde{t}'})_{4i}P_R\Big)\tilde{X}\tilde{\mathcal{U}}_i.\label{X'Stu}
\end{eqnarray}

Beyond the MSSM, there are couplings for baryon neutralino, quarks and squarks. They are deduced
in our previous work\cite{DCPC}, and can give new contributions to the quark EDMs.
\begin{eqnarray}
&&\mathcal{L}(\chi^0_Bq\tilde{q})
=\sum_{I,i=1}^3\sum_{j=1}^6 \frac{\sqrt{2}}{3}g_B\bar{\chi}_{B_i^0}(Z_{N_B}^{1i}Z_{\tilde{U}}^{Ij*}P_L-Z_{N_B}^{1i*}Z_{\tilde{U}}^{(I+3)j*}P_R)u^I\tilde{U}_j^*
+H.c.
\end{eqnarray}

\section{Formulation}

 Using the effective Lagrangian\cite{neuEDM} method, one obtains the fermion EDM $d_{f}$ from
\begin{eqnarray}
&&{\cal L}_{EDM}=-{i\over2}d_{f}\overline{f}\sigma^{\mu\nu}\gamma_5
fF_{\mu\nu},
\label{eq1}
\end{eqnarray}
with $F_{\mu\nu}$ representing the
electromagnetic field strength, $f$ denoting a fermion field.
It is obviously that this effective Lagrangian is CP-violating.
In the fundamental interactions, this CP-violating Lagrangian
can not be obtained at tree level.
Considering the CP-violating electroweak theory,
one can get this effective Lagrangian from the loop diagrams.
The chromoelectric dipole moment (CEDM)
$\overline{f}T^a\sigma^{\mu\nu}\gamma_5 fG^a_{\mu\nu}$ of quark can also give contribution
to the quark EDM. $G^a_{\mu\nu}$ denotes the gluon field strength.

To describe the CP-violating operators obtained from loop diagrams,
the effective method is convenient. The coefficients of the quark EDM and CEDM
at the matching scale $\mu$ should be evolved down to the quark mass scale with the
renormalization group equations. At matching scale, we can obtain the effective Lagrangian with the CP-violating
operators. The effective Lagrangian containing operators relating with the quark EDM and CEDM are
\begin{eqnarray}
&&\hspace{2.0cm}{\cal L}_{_{eff}}=\sum\limits_{i}^4C_{_i}(\Lambda){\cal O}_{_i}(\Lambda)\;,
\nonumber\\
&&{\cal O}_{_1}=\overline{q}\sigma^{\mu\nu}P_{_L}qF_{_{\mu\nu}}
\;,~~~~~~~~~~{\cal O}_{_2}=\overline{q}\sigma^{\mu\nu}P_{_R}qF_{_{\mu\nu}}
\;,\nonumber\\
&&{\cal O}_{_3}=\overline{q}T^a\sigma^{\mu\nu}P_{_L}qG^a_{_{\mu\nu}}
\;,~~~~~~{\cal O}_{_4}=\overline{q}T^a\sigma^{\mu\nu}P_{_R}qG^a_{_{\mu\nu}}
\;.
\label{eq3}
\end{eqnarray}
with $\Lambda$ representing the energy scale, where the Wilson coefficients $C_{i}(\Lambda)$ are
evaluated.

In our previous work\cite{weBLCPV},  we have studied the neutron EDM in the CP-violating BLMSSM, where
the contributions from baryon neutralino-squark and $\tilde{X}$-exotic squark are neglected, because they are
all small in the used parameter space. Here we take into account all the contributions at one loop level to study
the c and t EDMs.
In the CP-violating BLMSSM, the one-loop corrections to the
quark EDMs and CEDMs can be divided into six types
according to the quark self-energy diagrams.
We divide the quark self-energy diagrams according to the virtual particles as:
1 gluino-squark, 2 neutralino-squark, 3 chargino-squark, 4 X-exotic quark,
5 baryon neutralino-squark, 6 $\tilde{X}$-exotic squark.

From the quark selfenergy diagrams, one obtains the needed triangle diagrams
 by attaching a photon or gluon on the internal lines in all possible ways.
After the calculation, we obtain the effective Lagrangian contributing to the
quark EDMs and CEDMs.
The BLMSSM is larger than MSSM and includes the MSSM contributions.
In Fig.(\ref{OLmssm}), we plot all the
 one loop self energy diagrams of the up-type quark.

\begin{figure}[h]
\setlength{\unitlength}{1mm}
\centering
\includegraphics[width=4.5in]{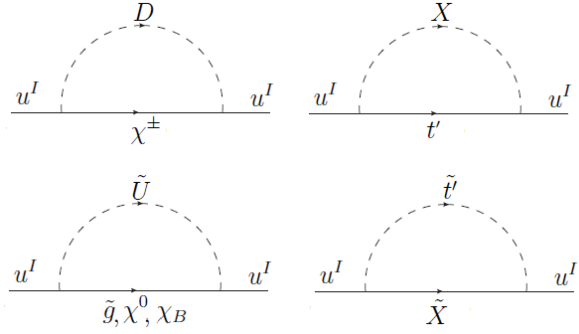}
\vspace{-10.0cm}
\caption[]{In the BLMSSM, one loop self energy diagrams are collected here,
and the corresponding triangle diagrams are obtained from them by attaching
a photon or a gluon in all possible ways.}\label{OLmssm}
\end{figure}

In this section, we show the one loop corrections to the
quark EDMs (CEDMs). The one loop chargino-squark contributions are
\begin{eqnarray}
&&d_{\chi_{k}^\pm}^\gamma(u^I)={e\alpha\over4\pi s_{W}^2}
V_{UD}^\dagger V_{DU}
\sum\limits_{i}^6\sum\limits_{k}^2{\bf Im}\Big((A_{C}^D)_{k,i}(B_{C}^D)^\dagger
_{i,k}\Big){m_{\chi_{k}^\pm}\over m_{\tilde{D}_i}^2}
\nonumber\\
&&\hspace{1.2cm}\times
\Big[-\frac{1}{3}\mathcal{B}\Big({m_{\chi_{k}^\pm}^2\over m_{\tilde{D}_i}^2}\Big)
+\mathcal{A}\Big({m_{\chi_{k}^\pm}^2\over
m_{\tilde{D}_i}^2}\Big)\Big],
\nonumber\\&&
d_{\chi_{k}^\pm}^g(u^I)={g_3\alpha\over4\pi s_{W}^2}
V_{UD}^\dagger V_{DU}
\sum\limits_{i}^6\sum\limits_{k}^2{\bf Im}\Big((A_{C}^D)_{k,i}(B_{C}^D)^\dagger
_{i,k}\Big){m_{\chi_{k}^\pm}\over m_{\tilde{D}_i}^2}\mathcal{B}\Big({m_{\chi_{k}^\pm}^2\over m_{\tilde{D}_i}^2}\Big),
\nonumber\\&&
(A_{C}^D)_{k,i}=\frac{m_{u^I}}{\sqrt{2}m_Ws_\beta}(Z_{\tilde{D}})^{Ji}(Z_+)^{2k},\nonumber\\&&
(B_{C}^D)_{k,i}=\frac{m_{d^I}}{\sqrt{2}m_Wc_\beta}(Z_{\tilde{D}})^{(J+3)i}(Z_-)^{2k}-
(Z_{\tilde{D}})^{Ji}(Z_-)^{1k}.\label{CHSD}
\end{eqnarray}
Here $\alpha=e^2/(4\pi),s_{W}=\sin\theta_{W},\;c_{W}=\cos\theta_{W}$,
$\theta_{W}$ is the Weinberg angle, $V$ is the CKM matrix.
We define the one loop functions $\mathcal{A}(r)$ and $\mathcal{B}(r)$ as\cite{xiangxiao}
\begin{eqnarray}
&&\mathcal{A}(r)=[2(1-r)^{2}]^{-1}[3-r+2\ln r/(1-r)], \nonumber\\&&
 \mathcal{B}(r)=[2(r-1)^2]^{-1} [1+r+2r\ln r/(1-r)].
\end{eqnarray}
$m_{\tilde{D}_i}\;(i=1\dots6)$ are the squarks masses and
$m_{_{\chi_{_k}^0}}\;(k=1,\;2,\;3,\;4)$ denote the eigenvalues of
neutralino mass matrix. $Z_{\tilde{D}_i}$ is the rotation matrix to diagonalize the
mass squared matrix for the down type squark.
$Z_-$ and $Z_+$ are the rotation matrices to obtain the mass eigenstates of charginos.

We show the gluino-squark  corrections to the quark EDMs and CEDMs
\begin{eqnarray}
&&d_{\tilde{g}}^\gamma(u^I)=-{4\over9\pi}e\alpha_{s}\sum\limits_{i=1}^6
{\bf Im}\Big(({\cal Z}_{\tilde U})^{(I+3)i}
({\cal Z}_{\tilde U})^{Ii*}e^{-i\theta_{3}}\Big)
{|m_{\tilde g}|\over m_{\tilde{U}_i}^2}\mathcal{B}\Big({|m_{\tilde g}|^2
\over m_{\tilde{U}_i}^2}\Big),\nonumber\\
&&d_{\tilde{g}}^g(u^I)={g_3\alpha_{s}\over4\pi}\sum\limits_{i=1}^6
{\bf Im}\Big(({\cal Z}_{\tilde U})^{(I+3)i}
({\cal Z}_{\tilde U})^{Ii*}e^{-i\theta_{3}}\Big)
{|m_{\tilde g}|\over m_{\tilde{U}_i}^2}\mathcal{C}\Big({|m_{\tilde g}|^2
\over m_{\tilde{U}_i}^2}\Big),\label{GLSU}
\end{eqnarray}
with $\alpha_{s}=g_{3}^2/(4\pi)$. ${Z}_{\tilde U}$ is the matrix for the up type squarks,
with the definition
${\cal Z}_{\tilde U}^\dagger{\bf m}_{\tilde U}^2{\cal
Z}_{\tilde U} =diag(m_{{\tilde q}_1}^2,\dots,m_{{\tilde
q}_6}^2)$. The concrete form of the loop function $\mathcal{C}(r)$ is\cite{xiangxiao}
\begin{eqnarray}
\mathcal{C}(r)=[6(r-1)^2]^{-1}[10r-26-(2r-18)\ln r/(r-1)].
\end{eqnarray}

To check the functions $\mathcal{A}(r),\mathcal{B}(r)$ and $\mathcal{C}(r)$ in the Ref.\cite{xiangxiao}, we calculate the one loop
triangle diagrams using the effective Lagrangian method. In the calculation,
we use the approximation
\begin{eqnarray}
\frac{1}{(k+p)^2-m^2}=1-\frac{2k\cdot p+p^2}{k^2-m^2}+\frac{4(k\cdot p)^2}{(k^2-m^2)^2},
\end{eqnarray}
with k representing the loop integral momentum and p representing the external momentum.
It is reasonable because the internal particles are at the order of TeV, and the external quark
is lighter than TeV, even for t quark. The ratio $\frac{m_t^2}{1TeV^2}\sim0.03$ is small enough
to use the approximation formula.

  For the diagram that the photon is just attached on the internal charged Fermions, our result corresponding
  to  $\mathcal{A}(r)$ is the function $a[F,S]$
\begin{eqnarray}
&&a[F,S]=\frac{\Lambda_{NP}^2}{i\pi^2}\int dk^4 \frac{k^2}{(k^2-m_F^2)^3(k^2-m_S^2)}\nonumber\\&&
=-\frac{F^2+2 S^2 \log (F)-4 F S+3 S^2-2 S^2 \log (S)}{2 (F-S)^3},\nonumber\\&&
\end{eqnarray}
with the definition $F=\frac{m_F^2}{\Lambda_{NP}^2}$ and $S=\frac{m_S^2}{\Lambda_{NP}^2}$.
$\lambda_{NP}$ represents the the energy scale of the new physics.
In order to compare with the function $\mathcal{A}(r)$, we use
$\lambda_{NP}^2=m_S^2, S\rightarrow 1, F\rightarrow\frac{m_F^2}{m_S^2}=r$ and obtain
\begin{eqnarray}
a[r,1]=-\frac{r^2-4 r+2 \log (r)+3}{2 (r-1)^3}=\mathcal{A}(r).
\end{eqnarray}

When the photon is just emitted from the internal charged scalars, our result for $\mathcal{B}(r)$ is
\begin{eqnarray}
&&b[F,S]=\frac{\Lambda_{NP}^2}{i\pi^2}\int dk^4 \frac{m_S^2}{(k^2-m_F^2)(k^2-m_S^2)^3}\nonumber\\&&
=\frac{F^2-2 F S \log (F)+2 F S \log (S)-S^2}{2 (F-S)^3},
\end{eqnarray}
With the same approach as that of $a[F,S]$, $b[F,S]$ turns into the form
\begin{eqnarray}
&&b[r,1]=\frac{r^2-2 r \log (r)-1}{2 (r-1)^3}=\mathcal{B}(r).
\end{eqnarray}

$\mathcal{C}(r)$ is obtained from the diagrams that the photon is attached on both the
internal charged Fermions and charged scalars.  Therefore,  $\mathcal{C}(r)$ is the
linear combination of  $\mathcal{A}(r)$  and $\mathcal{B}(r)$,
\begin{eqnarray}
&&\mathcal{C}(r)=\frac{1}{3}\mathcal{B}(r)-3\mathcal{A}(r).
\end{eqnarray}
From the above discussion, the results in Ref.\cite{xiangxiao} are the same with our results.
In our calculation, we do not ignore the mass of the
external fermion.
So, it is clear that the analytical expressions for the quark EDM in this work are practicable for both c and t.

Similarly, the contributions from the one loop neutralino-squark diagrams
 are also obtained
\begin{eqnarray}
&&
d_{\chi_{k}^0}^\gamma(u^I)={e\alpha\over12\pi s_{W}^2
c_{W}^2}\sum\limits_{i=1}^6\sum\limits_{k=1}^4{\bf Im}\Big((A_N)_{k,i}
(B_{N})^\dagger_{i,k}\Big)
{m_{\chi_{k}^ 0}\over m_{\tilde{U}_i}^2}\mathcal{B}\Big({m_{\chi_{k}^0}^2
\over m_{\tilde{U}_i}^2}\Big),
\nonumber\\&&
d_{\chi_{k}^0}^g(u^I)={g_3\alpha\over8\pi s_{W}^2
c_{W}^2}\sum\limits_{i=1}^6\sum\limits_{k=1}^4{\bf Im}\Big((A_N)_{k,i}
(B_{N})^\dagger_{i,k}\Big)
{m_{\chi_{k}^ 0}\over m_{\tilde{U}_i}^2}\mathcal{B}\Big({m_{\chi_{k}^0}^2
\over m_{\tilde{U}_i}^2}\Big),
\nonumber\\
&&(A_N)_{k,i}=-\frac{4}{3}s_W(Z_{\tilde{U}})^{(I+3)i}(Z_N)^{1k}+\frac{m_{u^I}c_W}{m_Ws_\beta}(Z_{\tilde{U}})^{Ii}(Z_N)^{4k},
\nonumber\\&&
(B_N)_{k,i}=(Z_{\tilde{U}})^{Ii}(\frac{s_W}{3}(Z_N)^{1k*}+c_W(Z_N)^{2k*})
+\frac{m_{u^I}c_W}{m_Ws_\beta}(Z_{\tilde{U}})^{(I+3)i}(Z_N)^{4k*}. \label{NESU}
\end{eqnarray}
$Z_N$ is the mixing matrix to get the eigenvalues $m_{\chi_k^0}$ (k = 1, 2, 3, 4) of
neutralino mass matrix. In the MSSM, there are also the front three type contributions Eqs.(\ref{CHSD})(\ref{GLSU})(\ref{NESU}).

At one loop level, there are three new type corrections beyond MSSM. The corrections from the virtual X and exotic up-type quark has
been deduced in the work\cite{weBLCPV}
\begin{eqnarray}
&&d^{\gamma}_{X_j}(u^I)=\frac{e\lambda_1\lambda_2}{24\pi^2}\sum_{i,j=1}^2
\frac{m_{t_{i+3}}}{m_{X_{j}}^2}{\bf Im}\Big((W_t)_{1i}(Z_X)^*_{1j}(U_t)_{2i}^*(Z_X)_{2j}\Big)
   \mathcal{A}\Big(\frac{m_{t_{i+3}}^2}{m_{X_{j}}^2}\Big),
\nonumber\\
&&
d^{g}_{X_j}(u^I)=\frac{g_3\lambda_1\lambda_2}{16\pi^2}\sum_{i,j=1}^2
\frac{m_{t_{i+3}}}{m_{X_{j}}^2}{\bf Im}\Big((W_t)_{1i}(Z_X)^*_{1j}(U_t)_{2i}^*(Z_X)_{2j}\Big)
   \mathcal{A}\Big(\frac{m_{t_{i+3}}^2}{m_{X_{j}}^2}\Big),\label{XEQ}
\end{eqnarray}
$m_{t_{i+3}}$ and $m_{X_{i}}$(i=1,2) are mass eigenvalues of the exotic up type quarks and X superfields.
 $W_t, U_t$ and $Z_X$ are the mixing matrices
defined in the Eqs.(\ref{Xchang})(\ref{ExoticQ}).

The one loop baryon neutralino and up-type squark contributions read as
\begin{eqnarray}
&&d_{\chi_{B}}^\gamma(u^I)=-{eg_B^2\over108\pi^2}\sum\limits_{i=1}^2\sum\limits_{j=1}^6
{\bf Im}\Big((Z_{N_B}^{1i})^2Z_{\tilde{U}}^{(I+3)j}Z_{\tilde{U}}^{Ij*}\Big)
{m_{\chi_{B}^i}\over m_{\tilde{U}_j}^2}\mathcal{B}\Big({m_{\chi_{B}^i}^2
\over m_{\tilde{U}_j}^2}\Big),
\nonumber\\&&
d_{\chi_{B}}^g(u^I)=-{g_3g_B^2\over72\pi^2}\sum\limits_{i=1}^2\sum\limits_{j=1}^6
{\bf Im}\Big((Z_{N_B}^{1i})^2Z_{\tilde{U}}^{(I+3)j}Z_{\tilde{U}}^{Ij*}\Big)
{m_{\chi_{B}^i}\over m_{\tilde{U}_j}^2}\mathcal{B}\Big({m_{\chi_{B}^i}^2
\over m_{\tilde{U}_j}^2}\Big).
\end{eqnarray}
$m_{\chi_{B}^i}$ (i=1,2,3) are the eigenvalues of baryon neutralino masses.

The exotic up-type squark and $\tilde{X}$ can also contribute to the c(t) EDM and CEDM
\begin{eqnarray}
&& d_{\tilde{X}}^\gamma={e\lambda_1\lambda_2\over24\pi^2}\sum\limits_{i=1}^2{\bf Im}
\Big((Z_{\tilde{t}'})^{3i*}(Z_{\tilde{t}'})^{4i*}\Big)
{m_{\tilde{X}}\over m_{\tilde{t}_{i+3}}^2}\mathcal{B}\Big({m_{\tilde{X}}^2
\over m_{\tilde{t}_{i+3}}^2}\Big),
\nonumber\\&&
d_{\tilde{X}}^g={g_3\lambda_1\lambda_2\over16\pi^2}\sum\limits_{i=1}^2{\bf Im}
\Big((Z_{\tilde{t}'})^{3i*}(Z_{\tilde{t}'})^{4i*}\Big)
{m_{\tilde{X}}\over m_{\tilde{t}_{i+3}}^2}\mathcal{B}\Big({m_{\tilde{X}}^2
\over m_{\tilde{t}_{i+3}}^2}\Big),\label{XESQ}
\end{eqnarray}
with $m_{\tilde{X}}$ and $m_{\tilde{t}_{i+3}}$  denoting the masses of $\tilde{X}$ and
exotic up-type squark respectively.

Using the renormalization group equations\cite{rge}, we evolve the coefficients of the quark EDM and
CEDM at matching scale $\mu$ down to the quark(c, t) mass scale\cite{CZHPD}
\begin{eqnarray}
d_{_q}^\gamma(\Lambda_{_\chi})=1.53d_{_q}^\gamma(\Lambda)
,~~~~ d_{_q}^g(\Lambda_{_\chi})=3.4d_{_q}^g(\Lambda),
\label{eq14}
\end{eqnarray}
The quark CEDMs can contribute to the quark EDMs at low energy scale. Therefore,
they must be taken into account in the numerical calculation and the formula is\cite{EaddC}
\begin{eqnarray}
&&d_{c}=d_{c}^\gamma+{e\over4\pi}d_{c}^g.
\label{eq15}
\end{eqnarray}

\section{The numerical results}
Here, the results are studied numerically.
We take into account not only the experiment constraints from Higgs and neutrino, but also
our previous works in this model. From ATLAS collaboration, $m_{\tilde{g}}\geq 1460\;$GeV is
the updated bound on the gluino mass\cite{mgmass}.
The parameters are supposed as
\begin{eqnarray}
&&
m_1 = m_2 = A_{BQ} = A_{BU} = 1\;{\rm TeV},~B_X = 500\;{\rm GeV},~
m_D^2 =\delta_{ij}\;{\rm TeV^2},~(i,j=1,2,3),\nonumber\\&&
A_u = A_d = A'_u = A'_d = 500\;{\rm GeV},~~~
Y_{d_4} = Y_{d_5} = 0.7Y_b,~~~\lambda_Q = \lambda_u = 0.5,
\nonumber\\&&
m^2_{\tilde{Q}_4} = m^2_{\tilde{Q}_5} = m^2_{\tilde{U}_4} = m^2_{\tilde{U}_5} = 1\;{\rm TeV^2},
~~~B_4 = \frac{3}{2},~~~A_{u_4} = A_{u_5} = 500\;{\rm GeV}.
\end{eqnarray}

\subsection{c quark EDM}
For the c quark EDM, we use the following parameters as
\begin{eqnarray}
&&\tan\beta = 10,~~~
\mu = 800\;{\rm GeV},~~~m_{\tilde{g}} = 1600\;{\rm GeV},\nonumber\\&&\tan\beta_B = 2,~~~
v_{B_t} = 3\;{\rm TeV},~~~Y_{u_4} = Y_{u_5} = 0.7Y_t.
\end{eqnarray}

\begin{figure}[h]
\setlength{\unitlength}{1mm}
\centering
\includegraphics[width=4.0in]{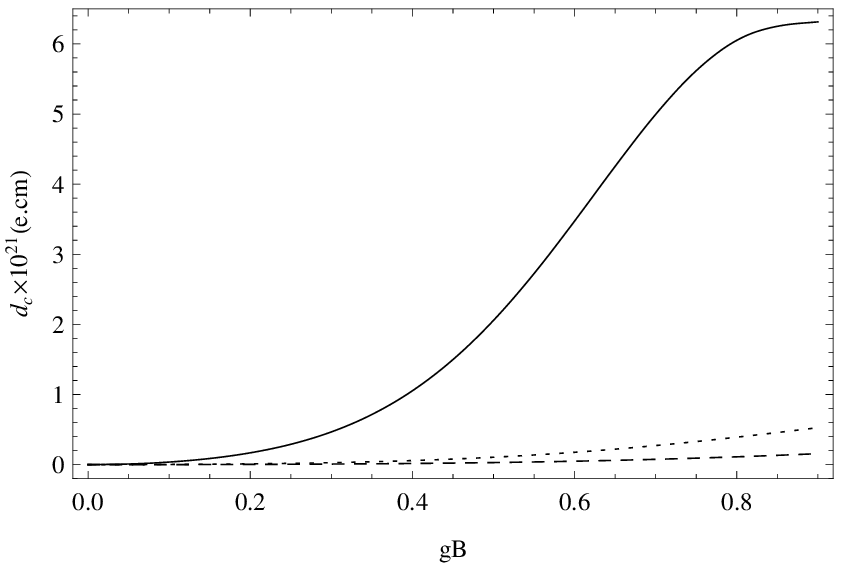}
\caption[]{The one loop corrections to c EDM versus gB with $\theta_{m_B}=-0.5\pi$,
the solid line, dotted line and dashed line corresponding to $m_B=(1,2,3)\times e^{{\rm i}\theta_{m_B}}{\rm TeV}$ respectively.}\label{GBCEDM}
\end{figure}

The baryon neutralino and squarks can give contributions to c quark EDM, which
is relevant to the parameters $g_B$ and $m_B$.  $m_B$ representing baryon gaugino masses, can have nonzero
CP-violating phase $\theta_{m_B}$. Both $m_B$ and $g_B$ influence the baryon neutralino masses. Furthermore, $g_B$
is the coupling constant for the quark-squark-baryon neutralino.
So, with $\theta_{m_B}=-0.5\pi,\; \lambda_1 = \lambda_2 = 0.1,~\mu_X = \mu_B = 3\;{\rm TeV},$
$m_Q^2=m_U^2 = \delta_{ij}{\rm TeV^2}~\texttt{for}~(i,j=1,2,3),$ we study c quark EDM versus $g_B$.
If we do not mention the other CP-violating phases, it indicates the other CP-violating phases are zero.
In Fig.\ref{GBCEDM}, the numerical results corresponding to $m_B=(1,2,3)\times e^{{\rm i}\theta_{m_B}}{\rm TeV}$
are plotted by the solid line, dotted line and dashed line respectively. In the whole, the three lines are
all increasing functions of $g_B$. The dotted line and the dashed line vary slightly. The solid line largen
quickly with the increasing $g_B$.  These three lines also imply the results are suppressed by large $|m_B|$.
The obtained numerical results from the nonzero CP-violating phase $\theta_{m_B}$ are at the order of $10^{-21}e.cm$,
which are four orders smaller than the upper bound of c quark EDM.

Here, we discuss the effects from the new phase $\theta_{\mu_B}$ of $\mu_B$, it can influence
the baryon neutralino masses. Based on the supposition $\mu_B = 2\times e^{{\rm i}\theta_{\mu_B}}\;{\rm TeV}\; (\theta_{\mu_B}=0.5\pi),\; \lambda_1 =\lambda_2 = 0.5,\; \mu_X = 3\;{\rm TeV}$, $m_Q^2= 4\delta_{ij}\;{\rm TeV^2},\;m_U^2= 2\delta_{ij}\;{\rm TeV^2}\;(i,j=1,2,3)$,
 the results versus $m_B$ are shown as the solid line for $g_B = \frac{1}{3}$.
 The solid line decreases quickly in the region $1000{\rm GeV}<m_B<1300{\rm GeV}$.
 When $m_B>1300{\rm GeV}$, the change extent of the solid line is small. The dotted line and the dashed line respectively represent the
 results for $g_B = \frac{1}{5}$ and $g_B = \frac{1}{10}$, and they are both the slowly
 decreasing functions of $m_B$. Generally speaking, the results are around $10^{-21}e.cm$ that are at the same order
 of Fig.\ref{GBCEDM}.

\begin{figure}[h]
\setlength{\unitlength}{1mm}
\centering
\includegraphics[width=4.0in]{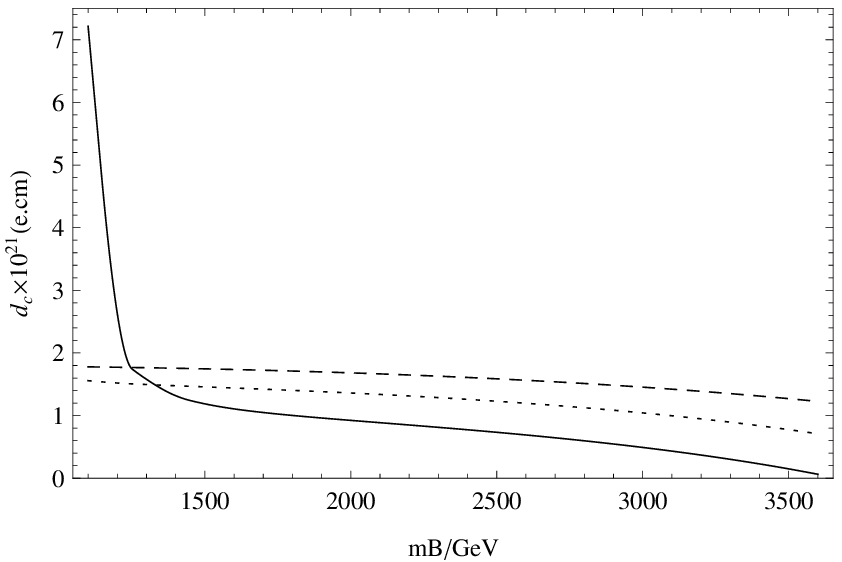}
\caption[]{The one loop corrections to c EDM versus mB with $\theta_{\mu_B}=0.5\pi$, the solid line,
dotted line and dashed line corresponding to $g_B=(1/3,\; 1/5,\;1/10)$ respectively.}\label{MBCEDM}
\end{figure}

\begin{figure}[h]
\setlength{\unitlength}{1mm}
\centering
\includegraphics[width=4.0in]{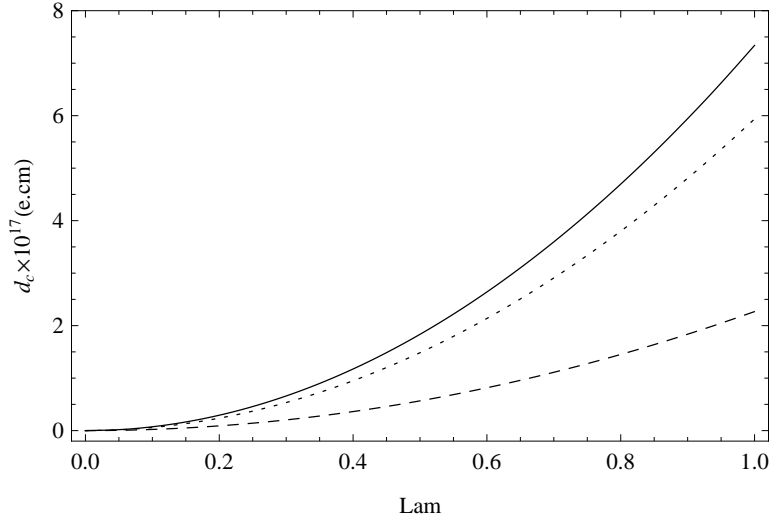}
\caption[]{The one loop corrections to c EDM versus Lam, while
the solid line, dotted line and dashed line denoting the results with
$\mu_X = e^{ i(0.5\pi,0.3\pi,0.1\pi)}\;{\rm TeV}$ respectively.}\label{LAMCEDM}
\end{figure}

$\lambda_1$ and $\lambda_2$ are important parameters for the couplings: quark-exotic quark-X and quark-exotic squark-$\tilde{X}$.
Therefore, the numerical results maybe be influenced obviously by the varying $\lambda_1$ and $\lambda_2$.
For simplicity, we suppose $\lambda_1 = \lambda_2 = Lam,\; g_B = \frac{1}{3},\;m_B = 1{\rm TeV}, \; \mu_B = 3{\rm TeV},\;m_Q^2=m_U^2= \delta_{ij}{\rm TeV^2}$ for $(i,j=1,2,3)$. With the nonzero
CP-violating phase $\theta_X=(0.5\pi, 0.3\pi, 0.1\pi)$ and $\mu_X = e^{ i\theta_X}\;{\rm TeV}$, the results versus $Lam$ are
denoted respectively by the solid line, dotted line and dashed line. The three lines are the increasing
functions of $Lam$ and the results are around $10^{-17}e.cm$. As $Lam>0.8$,  the solid line even exceeds its
the upper bound $5.0\times10^{-17}e.cm$. From the Figs.(\ref{GBCEDM}, \ref{MBCEDM}, \ref{LAMCEDM}), one can find the
effects from $\theta_X$ are much larger than the effects from $\theta_{\mu_B}$ and $\theta_{m_B}$.

\subsection{t quark EDM}
\begin{figure}[h]
\setlength{\unitlength}{1mm}
\centering
\includegraphics[width=4.0in]{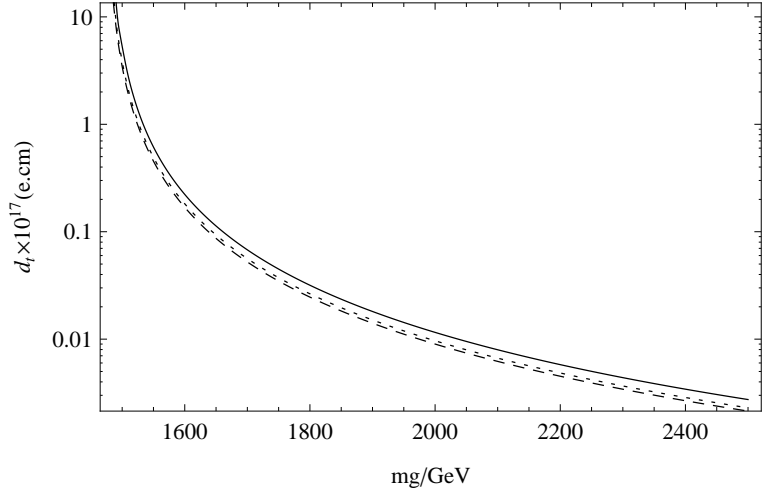}
\caption[]{The one loop corrections to t EDM versus $m_{\tilde{g}}$ with $\theta_3 = -0.5\pi$, the solid line,
dotted line and dashed line corresponding to $\tan\beta =(5,10,15)$.}\label{TBTEDM}
\end{figure}

To calculate the t quark EDM numerically, we use the parameters as
\begin{eqnarray}
&&g_B = \frac{1}{3},\;
m_B = 1\;{\rm TeV},\;
\mu_B = 3\;{\rm TeV}.
\end{eqnarray}
The quark-gluino-squark coupling corrections to t quark EDM are shown in Eq.(\ref{GLSU}).
$\tan\beta$ is important, because it influences the masses of chargino, neutralino, squark and so on.
The mixing matrices of squarks and exotic squarks have relation with $\tan\beta$.
The absolute value of gluino mass also influence the results obviously from Eq.(\ref{GLSU}).
 Using the parameters
$\theta_3 = -0.5\pi,\;\mu = 800\;{\rm GeV},\;
\tan\beta_B = 2,\;
Y_{u_4} = Y_{u_5} = 0.7Y_t,\;
\lambda_1 =\lambda_2 = 0.5,\;
\mu_X = 1\;{\rm TeV},\;V_{B_t} = 3300\;{\rm GeV},\;m_Q^2 =\delta_{ij}1500^2{\rm GeV^2},\;
 m_U^2 = \delta_{ij}{\rm TeV^2}$ with $(i,j=1,2,3)$, for t quark EDM we plot the results versus $m_{\tilde{g}}$.
In Fig.(\ref{TBTEDM}), the solid line ($\tan\beta=5$), dotted line ($\tan\beta=10$)
and dashed line ($\tan\beta=15$) are all decreasing functions of $m_{\tilde{g}}$.
During the $m_{\tilde{g}}$ region $(1500\sim2000)$ GeV, the results of the three lines shrink quickly.
Near the point $m_{\tilde{g}}=1480$ GeV, the theoretical predictions are at the order of $10^{-16}e.cm$ and
even reach $10^{-15}e.cm$. On the other hand, the extent of the influence from $\tan\beta$ is not large.

The effects to the t quark EDM from the $\mu$ parameter are also of interest. $\mu$
is included in the mass matrices of chargino and neutralino.
On the other hand $m_Q^2$ and $m_u^2$ can affect the masses and mixings of the up-type squark.
For simplification of the numerical discussion, we adopt the relation $m_Q^2=m_U^2=Mus^2$ and
use the parameters $\tan\beta = 10,\; m_{\tilde{g}} = 1600\;{\rm GeV},\;
 \tan\beta_B = 1.5,\; V_{B_t} = 3600\;{\rm GeV},\;Y_{u_4} = Y_{u_5} = 0.7Y_t,\;\lambda_1 = \lambda_2 = 0.1,\;
\mu_X = 1\;{\rm TeV}$. As $\theta_\mu=-0.5\pi$ and $\mu = (1500,\;2500,\;3500)e^{i\theta_\mu}\;{\rm GeV}$, the numerical results
for t quark EDM are plotted respectively by the dotted line, solid line and dashed line. The results are all decreasing
functions of $Mus$ and at the order of $10^{-18}e.cm$ as $Mus<3500\;{\rm GeV}$. The absolute value of $\mu$ also
influences the results, and the extent is small.

\begin{figure}[h]
\setlength{\unitlength}{1mm}
\centering
\includegraphics[width=4.0in]{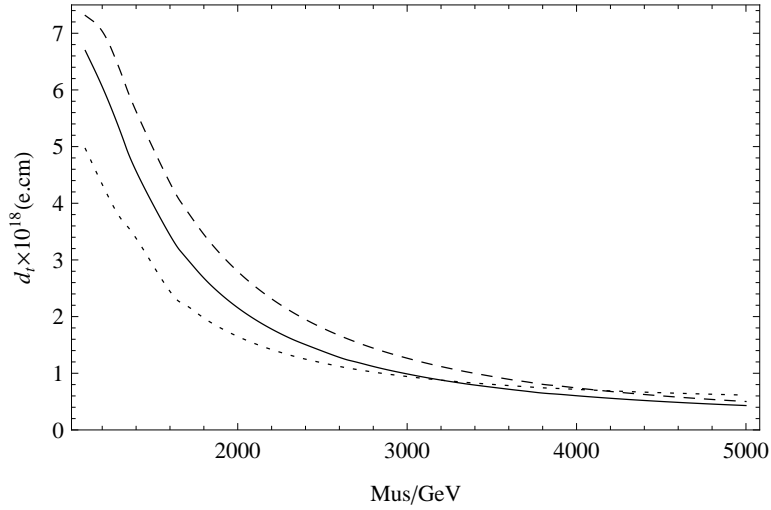}
\caption[]{The one loop corrections to t EDM versus Mus with $\theta_\mu=-0.5\pi$,
the dotted line, solid line and dashed line corresponding to the results with
$\mu = (1500,\;2500,\;3500)e^{i\theta_\mu}\;{\rm GeV}$.}\label{OLmssm}
\end{figure}

\begin{figure}[h]
\setlength{\unitlength}{1mm}
\centering
\includegraphics[width=4.0in]{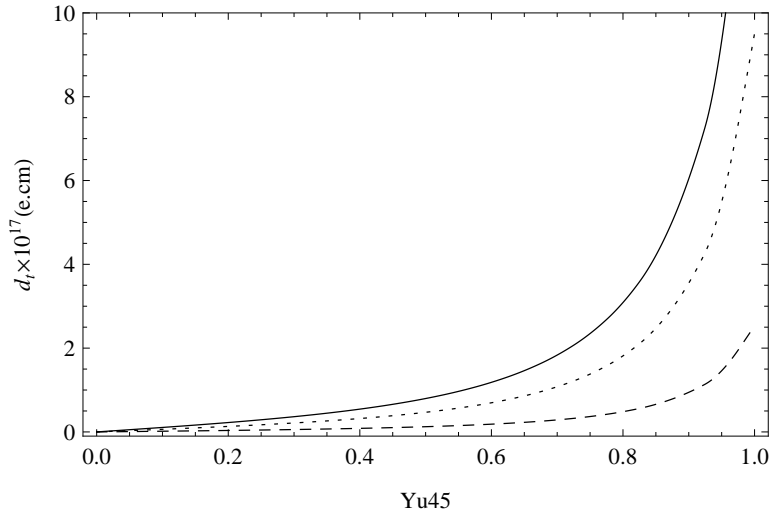}
\caption[]{The one loop corrections to t EDM versus Yu45 with $\mu_X =e^{{\rm i}\theta_X}\;{\rm TeV}$,
the solid line, dotted line and dashed line corresponding to the results with
$\theta_X=(0.5\pi,~0.2\pi,~0.05\pi)$. }\label{YUTEDM}
\end{figure}

The exotic squarks and exotic quarks are in connection with $Y_{u_4}$and $Y_{u_5}$, and the related contributions
are shown in Eqs.(\ref{XEQ},\ref{XESQ}). As discussed in the front subsection, nonzero $\theta_X$ can give large
contributions. In Fig.\ref{YUTEDM} we plot the solid line, dotted line and dashed line versus $Yu45$ with
$Y_{u_4}=Y_{u_5}=Yu45*Yt,\;\theta_X=(0.5\pi,\;0.2\pi,\;0.05\pi)$ and $\mu_X = 1\;e^{{\rm i}\theta_X}\;{\rm TeV}$.
The other used parameters are
$\tan\beta = 10,\;\mu = 800 {\rm GeV},\;m_{\tilde{g}} = 1600\;{\rm GeV},\;
\tan\beta_B = 2,\;V_{B_t} = 3\;{\rm TeV},\;\lambda_1 = \lambda_2 = 0.5,
\;m_Q^2 =m_U^2 = \delta_{ij}{\rm TeV^2}$ with $(i,j=1,2,3)$. They are all increasing functions of $Yu45$
and as $Yu45>0.8$ the numerical results largen quickly. At the point $Yu45=0.7$, the solid line and dotted line are
larger than $1\times10^{-17}e.cm$. The three lines are at the order of $10^{-18}e.cm$, when $Yu45$ is small.
For c and t quark EDMs, in our used parameter space the CP-violating phases $\theta_3$ and $\theta_X$ are important and can
provide large contributions.

\section{discussion}
   In the CP-violating BLMSSM, there are new
CP-violating phases $\theta_X,\;\theta_{\mu_B},\theta_{m_B}$ beyond MSSM.
For the c quark EDM, we consider the conditions $\theta_X\neq 0,\;\theta_{\mu_B}\neq0$ and $\theta_{m_B}\neq0$ respectively.
The results show $\theta_X$ can give large contributions, even reach the experiment upper bound($5\times10^{-17}e.cm$) for c EDM.
The effects produced from $\theta_{m_B}$ and $\theta_{\mu_B}$ are at the order of $10^{-21}e.cm$, which are much smaller
than those from $\theta_X$.
   For the t quark EDM the CP-violating phases $\theta_3,\;\theta_{\mu}$ and $\theta_X$ are studied.
In both $\theta_X\neq0$ and $\theta_3\neq0$ conditions, we find $d_t$ at the order of $10^{-17}e.cm$.
Especially for nonzero $\theta_3$ with $m_{\tilde{g}}$ near its lower bound, t quark EDM can reach $10^{-16}e.cm$ and
even larger.  They
are both larger than the results for $\theta_\mu\neq0$.

In BLMSSM, at one loop level there are three type contributions
(1. the virtual X and exotic up-type quark, 2. baryon neutralino and up-type squark, 3. exotic up-type squark and $\tilde{X}$)
to quark EDM beyond MSSM.
For the contributions beyond MSSM, to obtain large $d_c$ and $d_t$ the CP-violating phase $\theta_X$ should
be nonzero in our used parameter space.
With only $\theta_X\neq 0$, the nonzero contributions come from Eqs.(23)(25). In the
Fig.4 $d_c\sim10^{-17}e.cm$ and in the Fig.7 $d_t\sim10^{-17}e.cm$, the EDMs $d_c$ and $d_t$ are of the same order of magnitude.
In the other works,  the EDMs $d_c$ and $d_t$ should be of different order. What is the reason in this work?
The reason can be found from the couplings in Eqs.(\ref{Xt'u},\ref{X'Stu}).
The coupling constants $\lambda_1$ and $\lambda_2$ are important parameters. In our numerical calculation,
we adopt that the values of $\lambda_1(\lambda_2)$ for c quark are same with the values of $\lambda_1(\lambda_2)$ for t quark.
That is to say, for the up type quark generation 2 and generation 3, the adopted values of $\lambda_1(\lambda_2)$ are the same.

From these numerical results and our
previous work on neutron EDM, we find $\theta_3$ and $\theta_X$ are important CP-violating phases.
$\tan\beta,\;Y_{u_4},\;Y_{u_5},\;m_{\tilde{g}},\;m_Q^2,\;m_U^2,\;\lambda_1,\;\lambda_2,\;V_{B_t}$ are also important.
In the whole, our numerical results are large to be detected in the future. The work can confine the parameter
space in this model and possess meaning to the relevant experiments for c and t quark EDMs.

\begin{acknowledgments}

The work has been supported by the National Natural Science Foundation of China (NNSFC)
with Grant No. 11275036, No. 11535002, No. 11605037, No. 11647120, the Fund of Natural Science Foundation
of Hebei Province(A2011201118), Natural Science Fund of Hebei University
with Grant No. 2011JQ05, No. 2012-242. Hebei Key Lab of Optic-Electronic Information and Materials,
the midwest universities comprehensive strength promotion project.

\end{acknowledgments}


\vspace{2.0cm}
\begin{thebibliography}{99}
\bibitem{CPVKB} J.M. Christensen, J.W. Cronin, V.L. Fitch, R. Turlay, Phys. Rev. Lett. {\bf 13} (1964) 138;
 K. Abe, et al., [Belle Collaboration], Phys. Rev. Lett. {\bf 87} (2001) 091802; B. Aubert, et al.,
[BaBar Collaboration], Phys. Rev. Lett. {\bf89} (2002) 201802.

\bibitem{cp1}J. Ellis, S. Ferrara, and D.V. Nanopoulos, Phys. Lett.
B. {\bf 114} (1982) 231; J. Polchinski and M. B. Wise, {\it ibid.}
{\bf 125} (1983) 393; P. Nath, Phys. Rev. Lett. {\bf 66} (1991) 2565; Y. Kizukuri
and N. Oshimo, Phys. Rev. D. {\bf 46} (1992) 3025; {\bf 45} (1992) 1806.

\bibitem{SMTwoLoop} E. Shabalin, Sov. J. Nucl. Phys. {\bf28} (1978) 75;
F. Sala, JHEP {\bf1403} (2014) 061; I. Khriplovich, Phys. Lett. B {\bf173} (1986) 193-196;
A. Czarnecki and B. Krause, Phys. Rev. Lett. {\bf78} (1997) 4339-4342.

\bibitem{Higgs}CMS Collaboration, Phys. Lett. B {\bf 716} (2012) 30; ATLAS Collaboration, Phys. Lett. B {\bf 716} (2012) 1.

\bibitem{MSSM} J. Rosiek,  Phys. Rev. D {\bf 41} (1990) 3464; arXiv:hep-ph/9511250;
H.P. Nilles, Phys. Rept. {\bf 110} (1984) 1;
H.E. Haber and G.L. Kane, Phys. Rept. {\bf 117} (1985) 75-263.

\bibitem{munuSSM}N. Escudero, D.E. Lopez-Fogliani, C. Munoz and R.R. de Austri, JHEP {\bf12} (2008) 099;
Hai-Bin Zhang, Tai-Fu Feng, Fei Sun, et al., Phys. Rev. D {\bf89} (2014) 115007.
\bibitem{EDM}A. Pilaftsis, Phys. Rev. D {\bf 58} (1998) 096010;
M. Carena, J. Ellis, A. Pilaftsis and C.E.M. Wagner, Nucl. Phys. B {\bf 586} (2000) 92;
Tai-Fu Feng, Lin Sun, Xiu-Yi Yang, Nucl. Phys. B {\bf 800} (2008) 221;
Tai-Fu Feng, Lin Sun, Xiu-Yi Yang, Phys. Rev. D {\bf 77} (2008) 116008.

\bibitem{NEEDM} I. Altarev et al., Phys. Lett. B {\bf276} (1992) 242;
K. Smith et al., Phys. Lett. B {\bf234} (1990) 191; C. Patrignani et al. (Particle Data Group),
Chin. Phys. C {\bf40} (2016) 100001.

\bibitem{xiangxiao}T. Ibrahim and P. Nath, Phys. Rev. D {\bf 57} (1998) 478;
Erratum Phys. Rev. D {\bf 58} (1998) 019901; Phys.Lett. B {\bf 418} (1998) 98;
 Erratum: Phys.Lett. B {\bf 460} (1999) 498.

\bibitem{BLMSSM2}P.F. Perez, Phys. Lett. B {\bf711} (2012) 353; J.M. Arnold, P.F. Perez, B. Fornal,
 and S. Spinner, Phys. Rev. D {\bf 85} (2012) 115024; P.F. Perez, M.B. Wise. Phys. Rev. D {\bf 84} (2011) 055015.

\bibitem{BLMSSM1} P.F. Perez and M.B. Wise, JHEP {\bf 1108} (2011) 068; Phys. Rev. D {\bf 82} (2010) 011901.

\bibitem{weBLMSSM}Tai-Fu Feng, Shu-Min Zhao, Hai-Bin Zhang, Yin-Jie Zhang, Yu-Li Yan, Nucl.Phys. B {\bf 871} (2013) 223.

\bibitem{weBLNCP}S.M. Zhao, T.F. Feng, H.B. Zhang, et al., JHEP {\bf11} (2014) 119.
S.M. Zhao, T.F. Feng, H.B. Zhang, et al., Phys. Rev. D {\bf92} (2015) 115016.
Shu-Min Zhao, Tai-Fu Feng, Xing-Xing Dong, et al., Nucl. Phys. B {\bf910} (2016) 225¨C239.

\bibitem{weBLCPV}S.M. Zhao, T.F. Feng, B. Yan, et al., JHEP {\bf10} (2013) 020;
F. Sun, T.F. Feng, S.M. Zhao, et al., Nucl. Phys. B {\bf888} (2014) 30;
S.M. Zhao, T.F. Feng, X.J. Zhan, et al., JHEP {\bf07} (2015) 124.

\bibitem{TopEDM}A. Cordero-Cid, J.M. Hernandez, G. Tavares-Velasco, J.J. Toscano, J.Phys.G {\bf35} (2008) 025004.
\bibitem{HQEDM}A.G. Grozin, I.B. Khriplovich, A.S. Rudenko, Nucl. Phys. B {\bf821} (2009) 285-290.
\bibitem{HQEEEDM}A.E. Blinov, A.S. Rudenko, Nucl. Phys. Proc. Suppl. {\bf189} (2009) 257-259.
\bibitem{BFEDM}Xiao-Jun Bi, Tai-Fu Feng, Xue-Qian Li, et al., hep-ph/0412360.
\bibitem{OCEDM}Filippo Sala, JHEP {\bf1403} (2014) 061; Z.Z. Aydin, U. Erkarslan,  Phys. Rev. D {\bf67} (2003) 036006.
\bibitem{DCPC} Xing-Xing Dong, Shu-Min Zhao, Hai-Bin Zhang, et al., Chin. Phys. C {\bf40} (2016) 093103.
\bibitem{neuEDM} Tai-Fu Feng, Xue-Qian Li, Jukka Maalampi, Xin-min Zhang, Phys. Rev. D {\bf 71} (2005) 056005.
\bibitem{rge}R. Arnowitt, J. Lopez, and D. V. Nanopoulos, Phys. Rev. D {\bf 42} (1990) 2423;
R. Arnowitt, M. Duff and K. Stelle, {\it ibid.} {\bf 43} (1991) 3085.
\bibitem{CZHPD}R. Arnowitt, J. Lopez, and D. V. Nanopoulos, Phys. Rev. D. {\bf42} (1990) 2423; R. Arnowitt,
M. Duff and K. Stelle, ibid. 43, 3085(1991).
\bibitem{EaddC} A. Manohar, and H. Georgi, Nucl. Phys. B. {\bf234} (1984) 189.
\bibitem{mgmass} M. Aaboud et.al,ATLAS Collaboration, Phys. Rev. D {\bf 94} (2016) 052009.
\end{thebibliography}
\end{document}